\documentclass[12pt]{article}
\pdfoutput=1
\usepackage{epsfig,amsfonts,amsthm}
\usepackage{amsmath,amssymb}
\usepackage[normalem]{ulem}
\usepackage{color}
\usepackage{array}
\newcommand{\be}{\begin{equation}}
\newcommand{\ee}{\end{equation}}
\newcommand{\bea}{\begin{eqnarray}}
\newcommand{\eea}{\end{eqnarray}}

\renewcommand{\Re}{\mathrm{Re }}
\renewcommand{\Im}{\mathrm{Im }}

\def\lsim{\mathrel{\rlap{\lower4pt\hbox{\hskip1pt$\sim$}}
    \raise1pt\hbox{$<$}}}         
\def\gsim{\mathrel{\rlap{\lower4pt\hbox{\hskip1pt$\sim$}}
    \raise1pt\hbox{$>$}}}         
    
\usepackage{amsmath}
\usepackage{amsfonts}
\usepackage{amssymb}
\usepackage{subfig}
\usepackage{wrapfig}
\usepackage{graphicx}

\def\beq{\begin{equation}}
\def\eeq{\end{equation}}
\def\bea{\begin{eqnarray}}
\def\eea{\end{eqnarray}}

\def\<{\left\langle}
\def\>{\right\rangle}

\newcommand{\bt}{\begin{tabular}}
\newcommand{\et}{\end{tabular}}

\usepackage{slashed}
\usepackage{amssymb} 
\usepackage{float}

\usepackage[justification=centering]{caption}

\usepackage{tikz}
\usetikzlibrary{decorations.pathmorphing,decorations.markings}

\usepackage{graphicx}

\tikzset{
photon/.style={decorate, decoration={snake,amplitude=2pt, segment length=5pt}, draw=black},
particle/.style={draw=black, postaction={decorate}, decoration={markings,mark=at position .5 with {\arrow[draw=black]{>}}}},
antiparticle/.style={draw=black, postaction={decorate}, decoration={markings,mark=at position .5 with {\arrow[draw=black]{>}}}},
gluon/.style={decorate, draw=black, decoration={coil,amplitude=4pt, segment length=5pt}}
goldstone/.style={draw=green,postaction={decorate},decoration={markings,mark=at position .5 with {\arrow[draw=blue]{>}}}}
}

\usepackage[
top    = 3.5cm,
bottom = 3.5cm,
left   = 3.5cm,
right  = 3.5cm]{geometry}

\allowdisplaybreaks[2]
\begin{document}
\bibliographystyle{OurBibTeX}

\title{\hfill ~\\[-30mm]
                  \textbf{Higgs--Flavon Mixing and $h \rightarrow \mu\tau$
                }        }
\author{\\[-5mm]
Katri Huitu\footnote{E-mail: {\tt Katri.Huitu@helsinki.fi}},\ 
Venus Keus\footnote{E-mail: {\tt Venus.Keus@helsinki.fi}},\ \\
Niko Koivunen\footnote{E-mail: {\tt Niko.Koivunen@helsinki.fi}},\ 
Oleg Lebedev\footnote{E-mail: {\tt Oleg.Lebedev@helsinki.fi}}
\\ \\
  \emph{\small Department of Physics and Helsinki Institute of Physics,}\\
  \emph{\small Gustaf H{\"a}llstr{\"o}min katu 2, FIN-00014 University of Helsinki, Finland}\\[4mm]}
\maketitle

\vspace*{-0.250truecm}
\begin{abstract}
\noindent
{ATLAS and CMS have reported an excess in the flavor violating decay of the Higgs boson,
$h \rightarrow   \mu \tau$. We show that this result can be accommodated
through a mixing of the Higgs with a flavon, the field  responsible for generating the Yukawa matrices in the lepton sector. We employ a   version of the Froggatt-Nielsen
mechanism at the electroweak scale, with only the leptons and the flavon transforming non--trivially under the corresponding symmetry group. Non--observation of charged lepton flavor violation (LFV) in other processes imposes important constraints on the model, which
we find to be satisfied in substantial regions of parameter space.} 
\end{abstract}
 
\thispagestyle{empty}
\vfill
\newpage
\setcounter{page}{1}

\section{Introduction}
In the Standard Model (SM), the fermion Yukawa couplings are free parameters with no explanation for the   hierarchy among the fermion masses spanning over six orders of magnitude. Several Beyond the Standard Model (BSM) scenarios have been proposed to resolve this puzzle. A popular BSM framework was suggested by Froggatt and Nielsen developing a mechanism that naturally generates the SM fermion Yukawa couplings \cite{Froggatt:1978nt}. 

Lepton Flavour Violating (LFV) processes are absent in the SM, which  has been consistent with observations. Yet, recently the CMS and ATLAS experiments have hinted at the existence of a flavour violating decay of the Higgs boson $h\to\mu\tau$ \cite{Khachatryan:2015kon,Aad:2015gha}.\footnote{The process $h\to l_i l_j$ includes both  $h\to l_i^{+}l_j^{-}$ and $h\to l_i^{-}l_j^{+}$.}   The combined branching ratio for this decay 
is found to be 
\begin{equation}
{\rm BR}(h\to\mu\tau) = 0.82^{+0.33}_{-0.32} \;\% \;,
\label{h-tau-mu}
\end{equation}
while it is zero in the SM.
In this paper, we study the possibility that  this observation is due to a mixing between the SM Higgs field and a $flavon$, which is an integral part of the Froggatt-Nielsen mechanism \cite{Froggatt:1978nt}. The latter requires the existence of a scalar field (flavon) charged under an extra $U(1)$-symmetry which is broken spontaneously by its  Vacuum Expectation Value (VEV). The usual Higgs--portal \cite{Patt:2006fw}  coupling between the Higgs and the flavon field then leads to the Higgs--flavon mixing or, in other words, the existence of two mass eigenstates $H_1$ and $H_2$.
The lighter state, $H_1$, is identified with the 125 GeV Higgs--like scalar $h$  observed at the LHC, while the heavier state, $H_2$, has a dominant flavon component. Both of these scalars possess flavor changing couplings due to a misalignment between the lepton mass matrix 
and the matrix of the scalar couplings (see also  \cite{Babu:1999me,Giudice:2008uua,Bauer:2015fxa}).

To avoid the appearance of a Goldstone boson, the   Froggatt-Nielsen symmetry should be gauged, discrete or softly broken. We find that the gauge symmetry option is strongly constrained  and does not lead to a substantial
   $ {\rm BR}(h\to\mu\tau)$. On the other hand, the discrete and softly broken versions of the Froggatt-Nielsen mechanism can be viable. In this work, we focus on the $leptophilic$ flavon which generates flavor structures in the lepton sector only,   while the quark sector may possess its own flavon(s). In this case, the quark flavor changing processes do not constrain our model.  

Lepton flavor violation induced by the Standard Model Higgs has been the subject of intense research in recent years, starting with Ref.~\cite{Goudelis:2011un} where it was found that the low energy LFV constraints in the $(\mu,\tau)$ sector are quite weak. Ref.~\cite{Blankenburg:2012ex} 
  observed that a large $h\to\mu\tau$  rate comparable to that of  $h\to\tau\tau$ is consistent with these bounds.  This idea received a boost from experiment  when
  the tentative signal (\ref{h-tau-mu}) was detected which was followed by a surge in theory constructions. Relevant analyses of Higgs--induced lepton flavor violation include Refs.~\cite{McKeen:2012av}-\cite{Alvarado:2016par}. Our approach here differs from previous work in a few aspects. In particular,  we treat leptons and quarks on a different basis, which allows for the electroweak scale flavon sector. We also observe that the LFV processes in the Froggatt--Nielsen framework  are subject to certain 
  natural  cancellations.

The remainder of this paper is organized as follows. In Section \ref{FN-review}, we introduce our Froggatt--Nielsen set--up. In Section \ref{CLFV-bounds},
we  choose a favorable Yukawa texture and study all the relevant LFV constraints. 
The Higgs decay into leptons is analyzed in Section \ref{decays}, where we also provide an example of the parameter region saturating the experimental result (\ref{h-tau-mu}).
We conclude in Section \ref{conclusion}.

\section{The Froggatt-Nielsen framework}
\label{FN-review}

The salient feature of the Froggatt--Nielsen mechanism is that the SM Yukawa interactions are generated through  higher dimensional operators consistent with some U(1) symmetry
and the resulting Yukawa matrix is given in terms of the VEV of the flavon $\Phi$, 
\begin{equation}
c_{ij} ~ { \Phi^{n_{ij}} \over \Lambda^{n_{ij} } }   \bar f_{L,i} f_{R,j} ~H  + {\rm h.c.} ~,
\label{operator}
\end{equation} 
where $c_{ij}$ are order one coefficients,  $\Lambda$ is the new physics scale and
$f_{L,R}$ are SM fermions.  Such operators are
obtained by integrating out heavy states at the scale $\Lambda$. When $\Phi$ develops a VEV,
\begin{equation}
\Phi = {1\over \sqrt{2}} (v_\phi + \phi)
\end{equation}
with $\phi=\Re\phi+ i \Im\phi$, the Yukawa couplings are given by 
\begin{equation}
 Y_{ij}= c_{ij} ~ \left( { v_\phi \over \sqrt{2} \Lambda }\right)^{n_{ij} } \equiv c_{ij}~ \epsilon^{n_{ij} } \;,
 \label{Y}
\end{equation}
where $\epsilon$ is a small parameter. 
The U(1) invariance of the operator (\ref{operator})
requires 
\be 
 n_{ij} =-\frac{1}{q_{\phi}}(q_{\bar L,i}+q_{R,j}+q_{h}) \;,
\label{flavon power}
\ee 
where $q_i$ are the charges identified in Table \ref{U(1) charges}. The main attractive feature of the Froggatt--Nielsen  mechanism is that  order one charges translate into a hierarchy among  
the Yukawa couplings, thereby  eliminating unnaturally small dimensionless parameters from the flavour sector.

\begin{table}[h!]
\begin{center}
\begin{tabular}{|c|c|c|c|c|}
\hline
Particle   &  $ f_{L,i}^c$ & $f_{R,i}$ & $H$ & $\Phi$\\
\hline 
$U(1)$ charge & $q_{\bar L,i}$ & $q_{R,i}$ & $q_{h}$ & $q_{\phi}$ \\
\hline
\end{tabular}
\end{center}
\caption{The $U(1)$ charges of SM fermions $f_{R,L}$, SM Higgs field $H$ and the flavon $\Phi$.}
\label{U(1) charges}
\end{table}

In addition to the SM Yukawa couplings, integrating out heavy particles 
induces further operators, e.g. 
\begin{equation}
      \left( {\Phi \over \Lambda} \right)^{l_{ij}}  ~  \bar f_{L,i} \skew5 \not \partial  f_{L,j} ~~,~~   \partial_\mu \left( {\Phi \over \Lambda} \right)^{l_{ij}}  ~  \bar f_{L,i}  \gamma^\mu  f_{L,j}, 
\label{extra-ops}
\end{equation}
where $l_{ij}$ is some combination of charges, 
and similarly for the right--handed fermions. Here it is understood that if $l_{ij}<0$,
the flavon is to be replaced by its complex conjugate, $\Phi^{l_{ij}} \rightarrow   (\Phi^*)^{-l_{ij}}  $.  
 For $\Phi/\Lambda \ll 1$,
the fermion kinetic terms can be diagonalized by a $\Phi$--dependent field redefinition,
which also induces operators of the second type in (\ref{extra-ops}).
Due to U(1)-invariance the Yukawa textures   (\ref{Y}) are not affected by this transformation, while the order one coefficients can change.\footnote{This may also induce higher order corrections involving powers of $\Phi^* \Phi/\Lambda^2$.} 
Further, in the new basis the interaction terms involving $\partial_\mu\Phi$ can be rewritten   as the Yukawa terms   using the fermion equations of motion.
Therefore, the  $\epsilon$--dependence of the $\phi$--couplings is not affected by such manipulations and  we shall focus entirely on the Yukawa operator  (\ref{operator}).

In this work, we will only consider the lepton sector.
This is sufficient 
if only the leptons and the flavon transform  under the (leptonic) U(1) symmetry, while
the quark sector enjoys a different symmetry group.
As the Higgs field develops a VEV, 
\be 
H = \left(\begin{array}{c}
0\\
\frac{v+h}{\sqrt{2}}
\end{array}\right), \qquad 
 \label{field-definition}
\ee
the effective interaction involving the leptons and no more than one physical scalar takes the form 
\bea
\label{L-eff}
\mathcal{L}_{eff}&\supset &
 \frac{v}{\sqrt{2}} ~ Y_{ij} ~ \left(      1+ {h\over v} +  n_{ij}~{\phi \over v_\phi} \right)   \bar{l}'_{L,i}l'_{R,j} \;.   
 \eea
Here, the prime in the lepton fields serves to distinguish the weak basis ($l'$)
from the mass eigenstate basis ($l$). We see that while the Higgs interactions have the same flavour structure as the Yukawa matrices, those of the flavon do not which leads to flavour changing vertices.

Redefining the left-handed and right-handed leptons, one can diagonalize the lepton mass matrix. With  
 \be 
Y_{\textrm{diag}}=U_{L}YU_{R}^{\dagger}~,
\ee
where $U_{L,R}$ are unitary matrices, we get the following interactions in the mass eigenstate basis:
\bea
\mathcal{L}_{eff}&\supset & 
 \bar{l}_{L} \; M_{\textrm{diag}} \;l_{R} +
 {h\over \sqrt{2}} ~ \bar{l}_{L} \; Y_{\textrm{diag}} \; l_{R}
 + {\phi \over \sqrt{2}} {v\over v_\phi} ~\bar{l}_{L} \;  \kappa  \; l_{R}
   + {\rm h.c.},  
   \label{LFVinteraction}
\eea
where $M_{\textrm{diag}} = Y_{\textrm{diag}} v/\sqrt{2}$ and the flavon 
vertex involves the matrix 
\begin{equation}
\kappa = U_{L} \; (Y\cdot n) \; U_{R}^{\dagger}~,
\end{equation}
with $(Y\cdot n)_{ij} \equiv Y_{ij} n_{ij} $. 
Since $n_{ij}=a_i +b_j$, the matrix $Y\cdot n$  
can be expressed in terms of a matrix product which allows us to write $\kappa$ in a
closed form. Setting 
\begin{equation}
q_\Phi=-1~~,~~q_h=0 ~
\end{equation}
throughout this paper, we have 
\be
\kappa_{ij}= y_{j}\sum_{k=1}^{3}q_{\bar L,k}(U_{L})_{ik}(U_{L})^{\ast}_{jk}
+y_{i}\sum_{k=1}^{3}q_{R,k}(U_{R})_{ik}(U_{R})^{\ast}_{jk} ~,
\label{kappa}
\ee
where $y_{i}$ are the Yukawa matrix eigenvalues. This is the source of lepton flavor violation in our model. Upon the Higgs--flavon mixing, such flavor changing couplings 
also appear in the interactions of the physical Higgs--like boson. 

It is important to note that the neutrino sector does not have a direct impact on our considerations. Indeed, the LFV couplings are due to the matrices $U_{L,R}$ which diagonalize the charged lepton mass matrix. Here we simply assume that realistic neutrino textures can be generated by some mechanism which depends on the nature of the right--handed neutrinos and their multiplicity.\footnote{The number of right--handed neutrinos
can in principle be very large \cite{Buchmuller:2007zd}.}

Let us now turn to the scalar sector of the model.
The U(1)--symmetric scalar potential is given by 
\be 
V(H,\Phi)=-\mu_h^2(H^\dagger H) + \lambda_h(H^\dagger H)^2 -\mu_\phi^2(\Phi^\dagger \Phi) + \lambda_\phi(\Phi^\dagger \Phi)^2 + \lambda_{h\phi} (H^\dagger H)(\Phi^\dagger \Phi)~.
\ee
At the minimum of the potential, both $H$ and $\Phi$ develop VEVs leading to a mixing
between $h$ and Re$\phi$ (assuming CP invariance),  
 \be 
\left(\begin{array}{c}
H_1\\
H_2
\end{array}\right) \equiv 
\left(\begin{array}{cc}
\cos\theta & \sin\theta\\
-\sin\theta & \cos\theta 
\end{array}\right) \left(\begin{array}{c}
h\\
\Re \phi
\end{array}\right) ~.
\ee
The explicit form of $\theta$ and the $H_{1,2}$ masses in terms of the parameters of the potential is not needed
for our purposes and can be found elsewhere (see e.g.  \cite{Lebedev:2011aq}). We take
the lighter boson $H_1$ to be the 125 GeV Higgs--like scalar observed at the LHC and 
 parametrize our results in terms of $\theta$ and $m_{H_2}$. These are constrained by the 
 collider data, most importantly, by the LHC   and the electroweak precision measurements as summarized in   \cite{Falkowski:2015iwa}.
 
Since the vacuum breaks the Froggatt--Nielsen symmetry, a global U(1) would result in a massless Goldstone boson which is    phenomenologically  unacceptable. There are a few ways to circumvent this problem. One may gauge the U(1), however we find that this option
does not lead to interesting phenomenology due to tight constraints on a flavor non-universal Z$^\prime$. More interesting possibilities include discretizing the U(1)$\;\rightarrow \;$Z$_N$ or
introducing a soft explicit breaking.  In the Z$_N$ case, one can add  to the Lagrangian the operator
\begin{equation}
{c \over \Lambda^{N-4}} \;  \Phi^N +  {\rm h.c.} \;,
\end{equation}
which generates $m_{{\rm Im} \phi}$ of order $\sqrt{c} v_\phi (v_\phi / \Lambda)^{N/2-2}$.
Models with large $N \sim 10$ then include a rather light pseudoscalar  (unless $c$ is large). In the case of soft explicit U(1) breaking, one includes \cite{Tsumura:2009yf}
\begin{equation}
   { \tilde m}^2 \Phi^2 +  {\rm h.c.}  \;,
\end{equation}
which induces $m_{{\rm Im} \phi} \sim \tilde m$. In more involved models, such a mass term can also result from a   VEV of another scalar which does not couple to the SM fermions and thus does not alter the Yukawa textures.
In what follows, we will be agnostic as to the origin of the ${\rm Im} \phi$ mass
and will simply parametrize our results in terms of $m_{{\rm Im} \phi}$.

Finally, it is important to keep in mind the limitations of our effective field theory approach. While tree level processes are well under control, the loop contributions in our framework  are   only  indicative in nature since they depend on the details of the UV completion. This issue can be addressed in specific models whereas here we provide the expected lower limit on loop induced LFV.

\begin{table}[h!]
\begin{center}
\begin{tabular}{|p{0.25cm}|l|c|}
\hline  
 & Observable & Present limit \\[1mm]
\hline
1 & BR$(\mu\to eee)$  & $1.0\times 10^{-12}$ \cite{Bellgardt:1987du} 
\\[2mm]
2 & BR$(\tau\to eee)$  & $3.0\times 10^{-8}$ \cite{Amhis:2012bh} 
\\[2mm]
3 & BR$(\tau\to \mu\mu\mu)$  & $2.0\times 10^{-8} $ \cite{Amhis:2012bh}
\\[2mm]
4 & BR$(\tau^{-}\to\mu^{-}e^{+}e^{-})$  & $1.7\times 10^{-8}$ \cite{Hayasaka:2010np}
\\[2mm]
5 & BR$(\tau^{-}\to e^{-}\mu^{+}\mu^{-})$ & $2.7\times 10^{-8}$ \cite{Hayasaka:2010np}
\\[2mm]
6 & BR$(\tau^{-}\to e^{+}\mu^{-}\mu^{-})$ & $1.7\times 10^{-8}$ \cite{Hayasaka:2010np}
\\[2mm]
7 & BR$(\tau^{-}\to \mu^{+}e^{-}e^{-}$) & $1.5\times 10^{-8}$ \cite{Hayasaka:2010np}
\\[2mm]
8 & BR$(\mu\to e\gamma)$ & $5.7\times 10^{-13}$ \cite{Adam:2013mnn} 
\\[2mm]
9 & BR$(\tau\to \mu\gamma)$  & $4.4\times 10^{-8}$ \cite{Amhis:2012bh}
\\[2mm]
10 & BR$(\tau\to e\gamma)$ & $3.3\times 10^{-8}$ \cite{Amhis:2012bh}
\\[2mm]
11 & CR$(\mu$-$e, Au)$ & $7.0\times 10^{-13}$ \cite{Bertl:2006up}
\\[1mm]
\hline 
\end{tabular}
\caption{Current experimental bounds on the branching ratios of three--body LFV decays, magnetic transitions   and the conversion rate of $\mu \to e$.}
\label{experimental-bounds}
\end{center}
\end{table}

\section{Bounds on flavon--induced lepton flavor violation}
\label{CLFV-bounds}

The flavon interaction  (\ref{LFVinteraction}) induces lepton flavor violating processes 
which are strongly constrained by experiment. In this Section, we derive the corresponding
limits on the flavon couplings  parametrized by
\begin{equation}
\tilde \kappa_{ij} = {1 \over \sqrt{2}} {v \over v_\phi} \kappa_{ij} \;,
\end{equation}
such that the flavon--lepton coupling is  $ \tilde \kappa_{ij} \; \bar{l}_{L,i}  \; l_{R,j} \phi+ $ h.c.
In our analysis, we use the current   bounds from 
 the three--body decay  $l_{i}\to l_{j}l_{k}l_{l}$, magnetic transition $l_{i}\to l_{j}\gamma$ and $\mu \to e$ conversion processes   presented in Table \ref{experimental-bounds}.

 Throughout the paper  we   use a specific Yukawa texture which, as we show later, induces interesting LFV effects in Higgs decay and accommodates the $h\to\mu\tau$ result. 
Omitting for simplicity possible CP phases, 
the charge assignment shown in  Table \ref{charges-table1} leads to 
\be
Y = \left(
\begin{array}{ccc}
 3.4 ~\epsilon^6 & -0.6 ~\epsilon^6 & 3.5~ \epsilon^7 \\
 5.4 ~\epsilon^4 & 6.1 ~\epsilon^4  & -3.1 ~\epsilon^5 \\
 0.5 ~\epsilon^2  & 0.5 ~\epsilon^2    & 7.3 ~\epsilon^3 \\
\end{array}
\right), \quad 
\tilde \kappa = \frac{v}{v_\phi}\left(
\begin{array}{ccc}
1\times 10^{-5}  & -1\times 10^{-6}  & -3\times 10^{-6}  \\
-2\times 10^{-5}  & 2\times 10^{-3} & 6\times 10^{-4}  \\
3\times 10^{-4} & -4\times 10^{-3} & 2\times 10^{-2}
\nonumber
\end{array}
\right),
\ee
which reproduces the correct lepton masses for $\epsilon=0.1$ and the shown proportionality  coefficients (their precise values are given in Appendix A). 
The key feature here is that the Yukawa matrix is far from diagonal, leading
to a large $\mu-\tau$ mixing. 
Other possible textures will be explored in our subsequent work.

\begin{table}[ht!]
\begin{center}
\begin{tabular}{|c|c|c|c|c|c|c|c|c|}
\hline
Particle   &  $e_L^c$ & $e_R$ & $\mu_L^c$ & $\mu_R$ & $\tau_L^c$ &$\tau_R$ & $H$ & $\phi$
\\
\hline 
Charge & 6 & 0 & 4 & 0 & 2 & 1 & 0 & -1\\
\hline
\end{tabular}
\end{center}
\vspace{-3mm}
\caption{U(1)/Z$_N$ charge assignment. }
\label{charges-table1}
\end{table}

In the discrete symmetry  case, Z$_N$ acts on a given field by multiplying it with 
 $e^{2 \pi  q_i i /N}$ where $N$ is the order of Z$_N$ and $q_i$ is the corresponding charge from Table \ref{charges-table1}. In the allowed couplings the charges add up to zero mod $N$. 
The Yukawa texture then  has the above form for $N\geq 14$.\footnote{For smaller $N$, one can get allowed couplings at lower order in $\epsilon$ by replacing $\Phi$ with $\Phi^*$.} 
 Our LFV results equally apply to this case as well.

Given the texture, we can now derive bounds on the flavon VEV and mass.
If our U(1) is gauged, the flavon VEV has to be very large and no interesting effects in Higgs decay are expected. This can be seen, for instance, from the corresponding Z$^\prime$ contribution to $\mu \rightarrow eee$. Since the gauge coupling factors cancel between the vertices and the propagator,  this process probes  $v_\phi$ directly:
\begin{equation}
{ \Gamma (\mu \rightarrow eee) \over \Gamma (\mu \rightarrow e \nu \bar \nu)  } \bigg\vert_{Z^\prime} \sim
{v^4 \over v_\phi^4} \; \sin^2\delta_{e \mu}  < {\cal O}(10^{-12})\;,
\end{equation} 
where $\delta_{e \mu}$ is the mixing angle appearing at the Z$^\prime \bar e \mu$--vertex. 
For the textures we consider, $v_\phi $ has to be at least ${\cal O}(10$ TeV), which
makes $\tilde \kappa$ negligibly small.

In what follows, we therefore focus on the global symmetry case, which implies in particular 
that Im$\phi$ is a physical degree of freedom.
The diagrams which contribute to the LFV observables depend in general on the 
Higgs--flavon mixing.
 We thus consider in detail two cases: (i) negligible Higgs--flavon mixing,
  (ii) substantial  Higgs--flavon mixing.

\subsection{Negligible Higgs--flavon mixing }

When the Higgs--flavon mixing is close to zero, all lepton flavor violation is due to the exchange of  Re$\phi$ and 
Im$\phi$, which are mass eigenstates. This limiting case is instructive to consider and easy
to generalize to a more interesting scenario with a non--zero mixing.
To make the discussion more transparent we decouple Im$\phi$ in this subsection 
($m_{{\rm Im} \phi} \rightarrow \infty$),
whereas in the more realistic case of a non--zero mixing we take it properly into account.

 We start by studying the 3--body decays $l_{i}\to l_{j}l_{j}l_{j}$ (Fig.~\ref{CLFV-processes-fig}, left). These receive contributions both at tree level and 1--loop. The latter, with the photon attached to $l_{j}^+l_{j}^-$,  can be important since  they involve a tau in the loop whose coupling is enhanced by the tau mass. 
 The total decay rate for a given process is $\Gamma_{\textrm{tot}}=\Gamma^{\rm tree}+\Gamma^{\rm 1-loop}$. Due to the different helicity structures, the tree and loop amplitudes do not interfere.   
 We find for the most important processes\footnote{These results agree with those of Ref.~\cite{Harnik:2012pb}.}  
 
\bea
\Gamma^{\rm tree}(l_i \to l_j l_j l_j )&=&\frac{m_{i}^{5}}{4096\pi^{3}}
\frac{(\lvert \widetilde{\kappa}_{j i}\rvert^{2} +\lvert \widetilde{\kappa}_{i j}\rvert^{2})\lvert\widetilde{\kappa}_{jj}\rvert^{2}}{m_{{\rm Re}\phi}^{4}} ~,
\\[2mm]
\Gamma^{\rm 1-loop}(\mu \to e e e )&=& \frac{ \alpha^2 m_{\mu}^{3}m_{\tau}^{2}}{3072\pi^{5}}\frac{\lvert \widetilde{\kappa}_{\tau e}\rvert^{2}\lvert \widetilde{\kappa}_{\mu\tau}\rvert^{2} +\lvert \widetilde{\kappa}_{e \tau}\rvert^{2}\lvert\widetilde{\kappa}_{\tau\mu}\rvert^{2}}{m_{{\rm Re}\phi}^{4}}
\biggl[\frac{3}{2}- \log\left(\frac{m_{{\rm Re}\phi}^{2}}{m_\tau^{2}}\right) \biggr]^2\biggl[\log\left(\frac{m_{\mu}^{2}}{m_e^{2}}\right)-\frac{11}{4} \biggr],
\nonumber\\
 \Gamma^{\rm 1-loop}(\tau \to \mu \mu \mu )&=&
\frac{ \alpha^2 m_{\tau}^{5}}{3072\pi^{5}}\frac{(\lvert \widetilde{\kappa}_{\tau \mu}\rvert^{2} +\lvert \widetilde{\kappa}_{\mu\tau}\rvert^{2})\lvert \widetilde{\kappa}_{\tau \tau}\rvert^{2}}{m_{{\rm Re}\phi}^{4}}
\biggl[\frac{4}{3}-\log\left(\frac{m_{{\rm Re}\phi}^{2}}{m_\tau^2}\right) \biggr]^2\biggl[\log\left(\frac{m_{\tau}^{2}}{m_\mu^{2}}\right)-\frac{11}{4} \biggr],
\nonumber
\eea
and analogously for $\Gamma^{\rm 1-loop}(\tau \to e e e )$.

\begin{minipage}{\linewidth}
\begin{figure}[H]
\begin{tikzpicture}[thick,scale=1.0]
\fill[black] (1.5,0) circle (0.06cm);
\draw (1.5,0) -- node[black,above,xshift=-0.1cm,yshift=0.0cm] {$\widetilde{\kappa}_{j i}$} (1.5,0.03);
\draw[particle] (0,0) -- node[black,above,xshift=-0.9cm,yshift=-0.25cm] {$l_{i}$} (1.5,0);
\draw[particle] (1.5,0) -- node[black,above,xshift=1.0cm,yshift=0.2cm] {$l_{j}$} (3,1.0);
\draw[particle] (3,-0.75) -- node[black,above,yshift=0.15cm,xshift=0.7cm] {$l_{k}$} (4,-0.2);
\draw[particle] (4,-1.3) -- node[black,above,yshift=-0.6cm,xshift=0.7cm] {$\bar{l}_{l}$} (3,-0.75);
\draw[dashed] (1.5,0) -- node[black,above,yshift=-0.55cm,xshift=-0.15cm] {$\phi$} (3,-0.75);
\fill[black] (3,-0.75) circle (0.06cm);
\draw[particle] (3,-0.75) -- node[black,above,yshift=-0.7cm,xshift=0.0cm] {$\widetilde{\kappa}_{l k}$} (3,-0.78);
\end{tikzpicture}
\hspace{5mm}
\begin{tikzpicture}[thick,scale=1.0]
\draw[particle] (0,0) -- node[black,above,sloped,yshift=0.0cm,xshift=0.0cm] {$\mu$} (1.25,0);
\draw[particle] (1.25,0) -- node[black,above,sloped,yshift=0.0cm,xshift=0.0cm] {$\tau$} (2,0);
\draw[particle] (2,0) -- node[black,above,sloped] {$\tau$} (2.75,0);
\draw[particle] (2.75,0) -- node[black,above,sloped] {$e$} (4,0);
\draw[decorate,decoration={snake,amplitude=2pt,segment length=5pt},green] (2,0) -- node[black,above,yshift=-0.2cm,xshift=0.2cm] {$\gamma$} (2,-2);
\draw[particle] (0,-2) -- node[black,above,sloped,yshift=0.0cm,xshift=0.0cm] {$N$} (2,-2);
\draw[particle] (2,-2) -- node[black,above,sloped,yshift=0.0cm,xshift=0.0cm] {$N$} (4,-2);
\draw[dashed]  (1.25,0) node[black,above,sloped,yshift=0.7cm,xshift=0.8cm] {$\phi$}  arc (180:0:0.75cm) ;
\end{tikzpicture}
\hspace{5mm}
\begin{tikzpicture}[thick,scale=1.0]
\fill[black] (1,0) circle (0.06cm);
\draw (0,0) -- node[black,above,yshift=-0.8cm,xshift=0.0cm] {$\widetilde{\kappa}_{k i}$} (2,0);
\draw[particle] (0,0) -- node[black,above,sloped,yshift=-0.3cm,xshift=-0.9cm] {$l_{i}$} (1,0);
\draw[particle] (1,0) -- node[black,above,sloped,yshift=-0.7cm,xshift=0.0cm] {$l_{k}$} (3,0);
\draw[particle] (3,0) -- node[black,above,sloped,yshift=-0.3cm,xshift=0.9cm] {$l_{j}$} (4,0);
\draw[decorate,decoration={snake,amplitude=2pt,segment length=5pt},green] (3.2,0.9) -- node[black,above,yshift=-0.4cm,xshift=0.5cm] {$\gamma$} (4.5,2.2);
\draw[dashed]  (1,0) node[black,above,sloped,yshift=0.95cm,xshift=1.05cm] {$\phi$}  arc (180:0:1cm) ;
\fill[black] (3,0) circle (0.06cm);
\draw (3,0)  node[black,above,yshift=-0.8cm,xshift=0.0cm] {$\widetilde{\kappa}_{j k}$} (3,0);
\end{tikzpicture}
\vspace{0.5cm}
\caption{The $l_{i}\to l_{j}l_{k}l_{l}$ (left), $\mu\leftrightarrow e$-conversion (center) and $l_{i}\to l_{j}\gamma$ (right) processes mediated by the   flavon  $\phi$.  The decay $l_{i}\to l_{j}l_{k}l_{l}$ also receives important contributions at one loop.}
\label{CLFV-processes-fig} 
\end{figure}
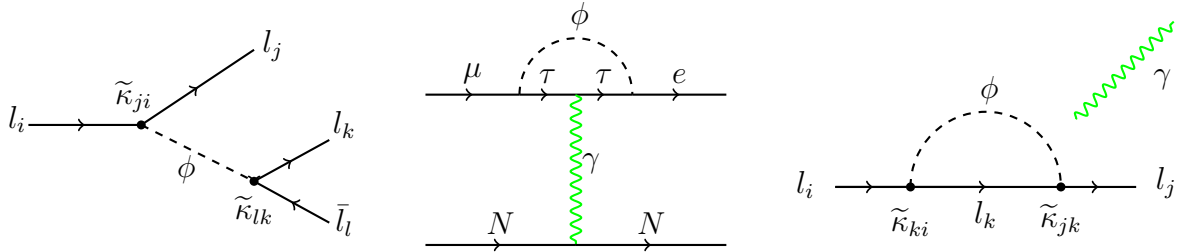    
\end{minipage}

\vspace{0.5cm}

The radiative transitions $l_{i}\to \l_{j}\gamma$ (Fig.~\ref{CLFV-processes-fig}, right) typically impose the strongest constraints on LFV models. In our case, the diagrams with the tau in the loop dominate. Neglecting the light lepton contributions, we find 

\begin{eqnarray}
\Gamma(\mu\to e\gamma)&=&
\frac{\alpha m_{\mu}^{3}m_{\tau}^{2}}{1024\pi^{4}} ~ \frac{|\widetilde{\kappa}_{e\tau}|^2 |\widetilde{\kappa}_{\tau\mu}|^2 + |\widetilde{\kappa}_{\tau e}|^2|\widetilde{\kappa}_{\mu\tau}|^2}{m_{{\rm Re} \phi}^{4}} \left[\frac{3}{2}-\log\left(\frac{ m_{{\rm Re} \phi}^{2}}{m_{\tau}^{2}}\right)\right]^{2},
\\
\Gamma(\tau\to \mu\gamma) &=&
\frac{\alpha m_{\tau}^{5}}{1024 \pi^{4}}~\frac{(|\widetilde{\kappa}_{\mu\tau}|^2+ |\widetilde{\kappa}_{\tau\mu}|^2)|\widetilde{\kappa}_{\tau\tau}|^2}{m_{{\rm Re} \phi}^{4}}\left[\frac{4}{3}-\log\left(\frac{m_{{\rm Re} \phi}^{2}}{m_{\tau}^{2}}\right)\right]^{2}.
\nonumber
\end{eqnarray}
An analogous expression holds for $\Gamma(\tau\to e\gamma)$ as well.
 
 \begin{table}[h!]
\begin{center}
\begin{tabular}{|l|c|l|}
\hline  
  Observable & Constraint\\[1mm]
\hline
  BR($\mu^- \to e^-e^-e^+$)  & \begin{footnotesize}
 $ \lvert \widetilde{\kappa}_{\tau e}\rvert ~ \lvert \widetilde{\kappa}_{\mu\tau}\rvert  
<
7.6 \times 10^{-6}$ \end{footnotesize}
\\[3mm]
  BR($\tau^- \to e^- e^- e^+$)  & \begin{footnotesize}$\lvert \widetilde{\kappa}_{\tau e}\rvert ~ \lvert \widetilde{\kappa}_{\tau \tau}\rvert 
 <
3.8 \times 10^{-3}$ \end{footnotesize}
\\[3mm]
  BR($\tau^- \to \mu^- \mu^- \mu^+$)  & \begin{footnotesize}$\lvert \widetilde{\kappa}_{\mu \tau}\rvert ~ 
 \lvert\widetilde{\kappa}_{\mu \mu }\rvert  
<
3.1 \times 10^{-3} $\end{footnotesize} 
\\[3mm]
  BR($\mu \to e \gamma$)  & \begin{footnotesize}
 $|\widetilde{\kappa}_{e\tau}|~ |\widetilde{\kappa}_{\tau\mu}| 
<
4.5\times 10^{-7}$ \end{footnotesize}
\\[3mm]
  BR($\tau \to \mu \gamma$)  & \begin{footnotesize}$ |\widetilde{\kappa}_{\tau\mu}|~| \widetilde{\kappa}_{\tau\tau}| 
<
4.9 \times 10^{-3}$ \end{footnotesize}
\\[3mm]
  BR($\tau \to e \gamma$)  & \begin{footnotesize}$|\widetilde{\kappa}_{\tau e}| ~|\widetilde{\kappa}_{\tau\tau}| 
<
4.2 \times 10^{-3}$\end{footnotesize}  
\\[1mm]
\hline 
\end{tabular}
\end{center}
\caption{Strongest bounds on the LFV couplings for a symmetric $\widetilde\kappa$--texture 
and $m_{{\rm Re} \phi}=500$ GeV. For other flavon masses, the bounds rescale approximately by $(m_{{\rm Re} \phi}/500\; {\rm GeV} )^2$. }
\label{magnetic-bounds1}
\end{table}
 
Finally, we also include constraints from the $\mu\leftrightarrow e$ conversion
(Fig.\ref{CLFV-processes-fig}, center).
Since the flavon does not couple to quarks, it is a loop process mediated by a tau.
 The conversion rate   is
\be
\Gamma(\mu\leftrightarrow e)=\left\lvert \frac{iD}{2  m_{\mu}}A_{\mu\to e\gamma}^{L}+\widetilde{g}_{LV}^{(p)}V^{(p)}\right\rvert^{2}+\left\lvert \frac{iD}{2  m_{\mu}}A_{\mu\to e\gamma}^{R}+\widetilde{g}_{RV}^{(p)}V^{(p)}\right\rvert^{2},
\ee
where we use $\Gamma_{\mbox{capture}~ Au}=13.07\times 10^{6} s^{-1}$, and $D$ and $V^{(p)}$ are the  overlap integrals for the nucleus in question. For gold, these integrals are \cite{Kitano:2002mt} $
D = 0.189$ and $V =0.0974$ in units of $m_\mu^{5/2}$. Here the same distribution is assumed for neutrons and protons in the nucleus \cite{Alonso:2012ji}. The Wilson coefficients are
\be
\widetilde{g}_{LV}^{(p)}=\frac{\alpha}{6\pi}\frac{\widetilde{\kappa}_{e\tau}\widetilde{\kappa}_{\mu\tau}^{\ast}}{ m_{{\rm Re} \phi}^{2}}\left[-\frac{11}{6}+\log\left(\frac{m_{{\rm Re} \phi}^{2}}{m_{\tau}^{2}}\right)\right],
\ee 
and $\widetilde{g}_{RV}^{(p)}$ is obtained from $\widetilde{g}_{LV}^{(p)}$ by replacing $\widetilde{\kappa}_{ij}$ with $\widetilde{\kappa}_{ji}^{\ast}$.
The invariant amplitude $A_{\mu \to e\gamma}^{L}$ is
\be
\label{invariant amplitude mu}
A_{\mu\to e\gamma}^{L}=-\frac{ie}{32\pi^{2}}\widetilde{\kappa}_{\tau e}^{\ast}\widetilde{\kappa}_{\mu\tau}^{\ast}\left[\frac{3}{2}-\log\left(\frac{m_{{\rm Re} \phi}^{2}}{m^{2}_{\tau}}\right)\right]\frac{m_{\tau}}{m_{{\rm Re} \phi}^{2}}~.
\ee
 The corresponding expression for $A^{R}$ is obtained by replacing $\widetilde{\kappa}_{ij}$ with $\widetilde{\kappa}_{ji}^*$.

To illustrate the strength of the constraints,
in Table \ref{magnetic-bounds1} we present a summary of the resulting bounds on $\tilde \kappa_{ij}$
assuming a $symmetric$ $\tilde \kappa_{ij}$--texture and $m_{{\rm Re} \phi}=500$ GeV.
For other flavon masses, the bounds rescale approximately by $(m_{{\rm Re} \phi}/500\; {\rm GeV} )^2$. The $\mu\leftrightarrow e$ conversion does not impose a significant bound
in this case.

\begin{figure}[h!]
\begin{center}
\includegraphics[scale=0.7]{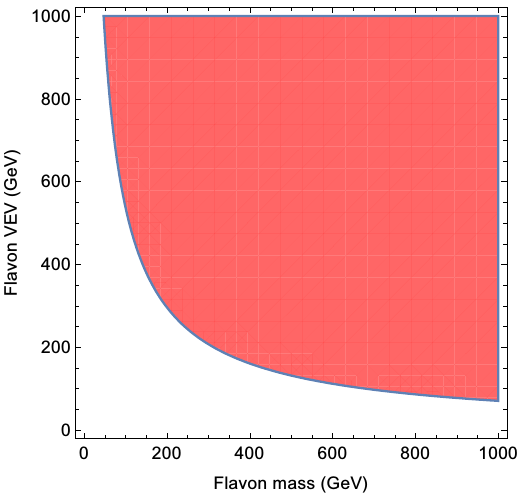}
\end{center}
\vspace{-3mm}
\caption{ Allowed parameter space (shaded) for the texture at hand (Eq.\ref{exact-texture}) with negligible Higgs-flavon mixing. The strongest constraint is imposed by BR($\mu \rightarrow e \gamma$).}
\label{Zeromixing}
\end{figure}

We find that the strongest constraint on the flavon mass and VEV for our texture is imposed by the $\mu \rightarrow e \gamma$ process. Figure \ref{Zeromixing} shows the allowed values of 
$v_\phi$ vs $m_{{\rm Re} \phi}$. As is clear from the above formulas, the bound scales approximately as $v_\phi \propto 1/m_{{\rm Re} \phi}$.  We see that a flavon VEV as low as 100 GeV is allowed if $m_{{\rm Re} \phi} \sim 500$ GeV. And conversely, a light flavon
with a mass smaller than  100 GeV is possible for $v_\phi > 1 $ TeV. 
This mass range is certainly consistent with collider constraints 
since the production cross section for a leptophilic flavon is highly suppressed.

Finally, since the couplings of Im$\phi$ are similar to those of Re$\phi$, analogous bounds apply to the mass of Im$\phi$.

\subsection{Substantial Higgs--flavon mixing }

The scalar  mass eigenstates mediating LFV are $H_1, H_2$ and Im$\phi$. The relevant interaction terms read
 \bea
\mathcal{L} &\supset& 
\left[\cos\theta~\frac{Y^{\rm diag}_{ij}}{\sqrt{2}}+\sin\theta~\widetilde{\kappa}_{ij}\right]
\bar{l}_{i}P_{R}l_{j} \;H_{1}
 +\left[-\sin\theta~\frac{Y^{\rm diag}_{ij}}{\sqrt{2}}+\cos\theta~\widetilde{\kappa}_{ij}\right]\bar{l}_{i}P_{R}l_{j} \; H_{2} 
 \nonumber\\
&+&
i  \widetilde{\kappa}_{ij} \; \bar{l}_{i}P_{R}l_{j}  \; \Im\phi 
+ {\rm h.c.}   
\label{couplings in mixing}
\eea
The couplings of $H_1, H_2$ to quarks are flavor--diagonal and   obtained 
by rescaling the corresponding SM couplings with $\cos\theta$ and  $-\sin\theta$, respectively.

Our previous tree and 1--loop level considerations can straightforwardly be generalized to the case at hand
 up to a trivial substitution of the lepton couplings  
and a summation over mass eigenstates. However, there are two significant changes. First, 
the $\mu \leftrightarrow e$ conversion  is now possible at tree level.  
Second, 
the important new ingredient is a set of 2--loop Barr-Zee diagrams \cite{Barr:1990vd}  with the top quark and the W in the loop (Fig.~\ref{2-loop}). Since both $H_1$ and $H_2$
have (flavor--diagonal) couplings to quarks, such diagrams   make a significant  
contribution to $\mu \rightarrow e \gamma$. 
 
 \begin{minipage}{\linewidth}
\centering
\begin{minipage}{\linewidth}
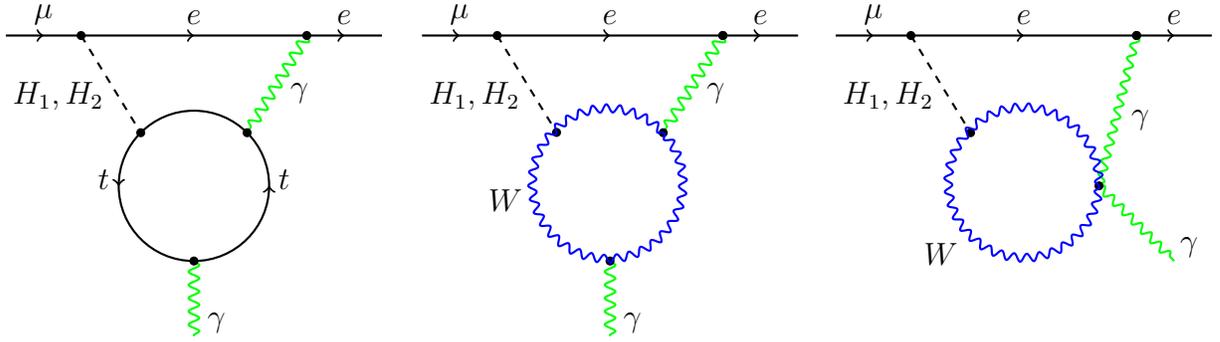
\begin{figure}[H]
\begin{tikzpicture}[thick,scale=1.0]
\draw[particle] (-0.5,0) -- node[black,above,sloped,yshift=-0.0cm,xshift=-0.0cm] {$\mu$} (0.5,0);
\fill[black] (0.5,0.0) circle (0.06cm);
\draw[particle] (0.5,0) -- node[black,above,sloped,yshift=-0.0cm,xshift=0.0cm] {$e$} (3.5,0);
\draw[particle] (3.5,0) -- node[black,above,sloped,yshift=-0.0cm,xshift=0.0cm] {$e$} (4.5,0);
\draw[dashed] (0.5,0) -- node[black,above,yshift=-0.5cm,xshift=-0.7cm] {$H_{1},H_2$} (1.293,-1.293);
\fill[black] (1.293,-1.293) circle (0.06cm);
\draw[decorate,decoration={snake,amplitude=2pt,segment length=5pt},green] (3.5,0) -- node[black,above,yshift=-0.4cm,xshift=0.3cm] {$\gamma$} (2.707,-1.293);
\fill[black] (2.707,-1.293) circle (0.06cm);
\draw[decorate,decoration={snake,amplitude=2pt,segment length=5pt},green] (2,-3) -- node[black,above,yshift=-0.6cm,xshift=0.3cm] {$\gamma$} (2,-4);
\fill[black] (2,-3) circle (0.06cm);
\draw[particle]  (2,-1) node[black,above,sloped,yshift=-1.2cm,xshift=-1.2cm] {$t$}  arc (90:270:1cm) ;
\draw[particle]  (2,-3) node[black,above,sloped,yshift=0.8cm,xshift=1.2cm] {$t$}  arc (-90:90:1cm) ;
\fill[black] (3.5,0) circle (0.06cm);
\end{tikzpicture}
\hspace{2mm}
\begin{tikzpicture}[thick,scale=1.0]
\draw[particle] (-0.5,0) -- node[black,above,sloped,yshift=-0.0cm,xshift=-0.0cm] {$\mu$} (0.5,0);
\fill[black] (0.5,0.0) circle (0.06cm);
\draw[particle] (0.5,0) -- node[black,above,sloped,yshift=-0.0cm,xshift=0.0cm] {$e$} (3.5,0);
\fill[black] (3.5,0) circle (0.06cm);
\draw[particle] (3.5,0) -- node[black,above,sloped,yshift=-0.0cm,xshift=0.0cm] {$e$} (4.5,0);
\draw[dashed] (0.5,0) -- node[black,above,yshift=-0.5cm,xshift=-0.7cm] {$H_{1},H_2$} (1.293,-1.293);
\fill[black] (1.289,-1.289) circle (0.06cm);
\draw[decorate,decoration={snake,amplitude=2pt,segment length=5pt},green] (3.5,0) -- node[black,above,yshift=-0.4cm,xshift=0.3cm] {$\gamma$} (2.707,-1.293);
\fill[black] (3.5,0) circle (0.06cm);
\fill[black] (2.707,-1.293) circle (0.06cm);
\draw[decorate,decoration={snake,amplitude=2pt,segment length=5pt},green] (2,-3) -- node[black,above,yshift=-0.6cm,xshift=0.3cm] {$\gamma$} (2,-4);
\fill[black] (2,-3) circle (0.06cm);
\draw[decorate,decoration={snake,amplitude=1.5pt,segment length=5pt},blue]  (2.707,-1.293) node[black,above,sloped,yshift=-1.2cm,xshift=-2.1cm] {$W$}  arc (45:405:1cm) ;
\hspace{5.5cm}
\draw[particle] (-0.5,0) -- node[black,above,sloped,yshift=-0.0cm,xshift=-0.0cm] {$\mu$} (0.5,0);
\fill[black] (0.5,0.0) circle (0.06cm);
\draw[particle] (0.5,0) -- node[black,above,sloped,yshift=-0.0cm,xshift=0.0cm] {$e$} (3.5,0);
\fill[black] (3.5,0) circle (0.06cm);
\draw[particle] (3.5,0) -- node[black,above,sloped,yshift=-0.0cm,xshift=0.0cm] {$e$} (4.5,0);
\draw[dashed] (0.5,0) -- node[black,above,yshift=-0.5cm,xshift=-0.7cm] {$H_{1},H_2$} (1.293,-1.293);
\fill[black] (1.293,-1.293) circle (0.06cm);
\draw[decorate,decoration={snake,amplitude=1.5pt,segment length=5pt},green] (3.5,0) -- node[black,above,yshift=-0.4cm,xshift=0.3cm] {$\gamma$} (3,-2);
\fill[black] (3.5,0) circle (0.06cm);
\draw[decorate,decoration={snake,amplitude=1.5pt,segment length=5pt},green] (3,-2) -- node[black,above,yshift=-0.6cm,xshift=0.7cm] {$\gamma$} (4,-3);
\fill[black] (3,-2) circle (0.06cm);
\draw[decorate,decoration={snake,amplitude=1.5pt,segment length=5pt},blue]  (3,-2) node[black,above,sloped,yshift=-1.2cm,xshift=-2.1cm] {$W$}  arc (0:360:1cm) ;
\end{tikzpicture}
\vspace{0.5cm}
\caption{  Barr--Zee diagrams  contributing to $\mu\to e\gamma$.  }
\label{2-loop}
\end{figure}
\end{minipage}
\end{minipage}
\vspace{0.5cm}

We find again  that the most important constraint on the flavon VEV for our texture comes from 
the  $\mu \rightarrow e\gamma$ process. Let us consider it in more detail. The relevant amplitude is a sum of the 1-- and 2--loop contributions,  
\be
A_{\mu\to e\gamma}^{L}=A_{\mu\to e\gamma}^{L}(1-\textrm{loop})+A_{\mu\to e\gamma}^{L}(2-\textrm{loop}) ~.
\ee
At one loop we have
\bea
A_{\mu\to e\gamma}^{L}(1-\textrm{loop})&=&
-\frac{ie m_{\tau}}{32\pi^{2}}\widetilde{\kappa}^{\ast}_{\tau e}\widetilde{\kappa}^{\ast}_{\mu\tau}
\biggl\{
\frac{\sin^{2}\theta}{m_{H_{1}}^{2}}\left[\frac{3}{2}-\log\left(\frac{m_{H_{1}}^{2}}{m^{2}_{\tau}}\right)\right]  \\
&&~~~~~~~~~~~~~~~
+\frac{\cos^{2}\theta}{m_{H_{2}}^{2}}\left[\frac{3}{2}-\log\left(\frac{m_{H_{2}}^{2}}{m^{2}_{\tau}}\right)\right]
-\frac{1}{m_{\Im\phi}^2}\left[\frac{3}{2}-\log\left(\frac{m_{\Im\phi}^{2}}{m^{2}_{\tau}}\right)\right]
\biggr\}.  \nonumber
\eea
The 2--loop amplitude receives contributions from the top quark and the $W$ boson \cite{Harnik:2012pb},
\be
A_{\mu\to e\gamma}^{L}(2-\textrm{loop})=A^{L}_{t}+A^{L}_{W},
\ee
with  
\be
A^{L}_{t}=-i\frac{e\alpha v G_{F} }{6\sqrt{2}\pi^3} \sin\theta\cos\theta \;\widetilde{\kappa}_{\mu e}^{\ast} \left[f(z_{tH_{1}})-f(z_{tH_{2}})\right]\;,
\ee
and
\bea
A^{L}_{W}&=&i\frac{e\alpha v G_{F}}{16\sqrt{2}\pi^3}\sin\theta\cos\theta \; \widetilde{\kappa}_{\mu e}^{\ast}\\
&&\times\left\{\left[3f(z_{WH_{1}})+5g(z_{WH_{1}})+\frac{3}{4}g(z_{WH_{1}})+\frac{3}{4}h(z_{WH_{1}})+\frac{f(z_{WH_{1}})-g(z_{WH_{1}})}{2z_{WH_{1}}} \right]\right.\nonumber\\
&&-\left.\left[3f(z_{WH_{2}})+5g(z_{WH_{2}})+\frac{3}{4}g(z_{WH_{2}})+\frac{3}{4}h(z_{WH_{2}})+\frac{f(z_{WH_{2}})-g(z_{WH_{2}})}{2z_{WH_{2}}} \right]\right\}.\nonumber
\eea
Here the loop functions are:
\bea
f(z)&=&\frac{1}{2}z\int_{0}^{1}dx\frac{1-2x(1-x)}{x(1-x)-z}\log\left(\frac{x(1-x)}{z}\right),\\
g(z)&=&\frac{1}{2}z\int_{0}^{1}dx\frac{1}{x(1-x)-z}\log\left(\frac{x(1-x)}{z}\right),\\
h(z)&=&z^{2}\frac{\partial}{\partial z}\left(\frac{g(z)}{z}\right)=\frac{z}{2}\int_{0}^{1}\frac{dx}{z-x(1-x)}\left[1+\frac{z}{z-x(1-x)}\log\left(\frac{x(1-x)}{z}\right)\right].
\eea
The arguments of these functions are defined by $z_{tH_{i}}=m_{t}^{2}/m^{2}_{H_{i}}$ and $z_{WH_{i}}=m_{W}^{2}/m^{2}_{H_{i}}$, with $i=1,2$. The $A_{\mu\to e\gamma}^{R}(2-\textrm{loop})$ amplitude is obtained by replacing  $\widetilde{\kappa}^{\ast}_{ji}$ with  $\widetilde{\kappa}_{ij}$. An analogous $Z$--boson contribution is suppressed compared to the photon one and therefore neglected.
The resulting $\mu\to e\gamma$  decay rate  is calculated according to 
\be
\Gamma(\mu\to e\gamma)=\frac{m_{\mu}^{3}}{4\pi}\left(\lvert A^L\rvert^{2}+\lvert A^R\rvert^{2}\right)\;.
\ee

\begin{figure}[ht!]
\begin{center}
\includegraphics[scale=0.571]{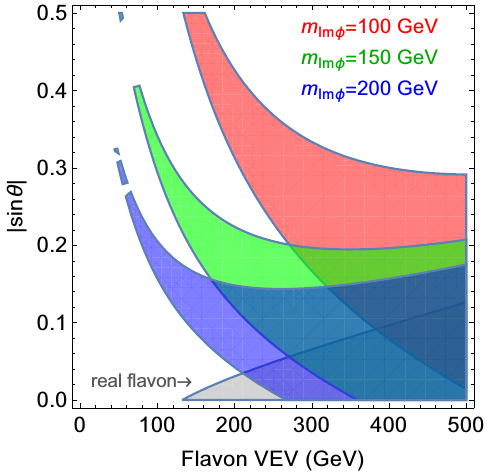}
\includegraphics[scale=0.745]{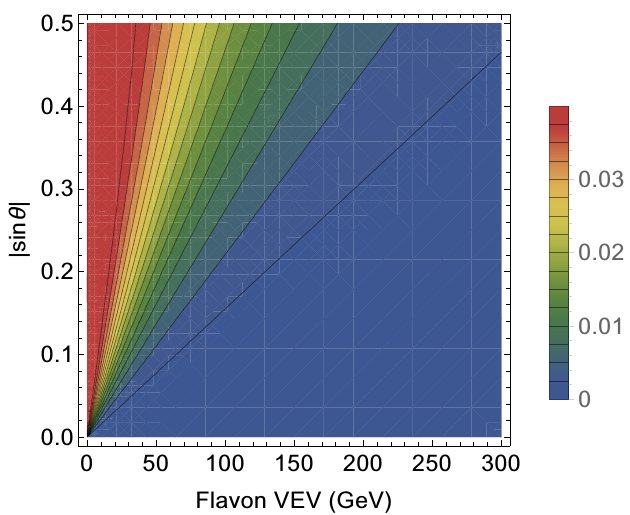}
\end{center}
\caption{Left:   parameter space allowed by the LFV constraints  for $m_{\Im\phi}=100,\;150, \;200$ GeV. We have set $m_{H_2}= 500$ GeV. (The discontinuities appear for technical reasons.) Right: BR$_{\rm eff} (H_{1}\to\mu\tau)$ as a function of $v_\phi$ and $\vert \sin\theta\vert$.
}
\label{1-LHC}
\end{figure}

Our results are presented in Fig.~\ref{1-LHC}  (left).
The shaded areas in the $(v_\phi, \sin\theta)$ plane    are allowed by all the LFV constraints (of which BR($\mu \rightarrow e \gamma$) is the strongest one) for a given $m_{H_2}$ and $m_{\Im\phi}$. Comparison to the real flavon case shows that considerable cancellations between the $H_i$ and Im$\phi$ contributions take place. 
These are due to  the pseudoscalar nature of Im$\phi$ which introduces a relative minus sign and the lightness of Im$\phi$ naturally expected in our framework. Similar cancellations apply also to the loop contribution for the $\mu \rightarrow eee$ process, while the $\mu \rightarrow e $ conversion bound is weaker even though it's not subject to the cancellations. 
   
We see that the flavon VEV is allowed to be as low as  100 GeV and a substantial Higgs--flavon mixing is consistent with the LFV data. The latter is also constrained by
  the collider and electroweak measurements as a function of
$m_{H_2}$ \cite{Falkowski:2015iwa}. 
In particular, $\vert \sin\theta\vert \simeq 0.3$
is allowed for $m_{H_2}=500$ GeV.  Fig.~\ref{1-LHC} shows that this value is  consistent 
with $v_\phi \sim 100$ GeV for a range of $m_{\Im\phi}$ around 150-200 GeV. As mentioned before, direct collider constraints on Im$\phi$ are very loose due to its   small couplings to leptons. Therefore, all of the relevant constraints are satisfied in that region. The right panel of  Fig.~\ref{1-LHC} then shows that one expects a substantial decay rate $H_{1}\to\mu\tau$, which we examine closely in the next section.

 \section{Leptonic Higgs decays  }
 \label{decays}
 
Our ultimate goal is to understand whether it is possible to obtain BR$(H_{1}\to\mu\tau)$ 
around 1\% in our simple leptonic Froggatt--Nielsen framework. The SM Higgs decay into tau's has a branching fraction of 6\%. Lepton flavor violating $H_1$ couplings are proportional to  $\vert \sin\theta\vert$ which cannot be greater than 0.35 or so,
resulting already  in an order of magnitude suppression. This makes it clear that  $\widetilde{\kappa}_{\mu\tau} $ and/or $\widetilde{\kappa}_{\tau\mu} $ must be  comparable to or larger than   the Higgs--tau Yukawa coupling in the Standard Model.
This can be achieved in the Froggatt--Nielsen framework,   yet it leads to the  
enhancement of the diagonal couplings as well. Since these are constrained by the LHC data, it is a non--trivial task to find a consistent model. One way to relieve the tension 
is to choose $\sin\theta<0$ which leads to some cancellations in $H_1 \rightarrow l_i l_i$ for our Yukawa texture. 

We find
  \bea
\Gamma(H_{1}\to\mu\tau)=
\frac{m_{H_{1}}}{8\pi}\sin^{2}\theta \; \left(\vert \widetilde{\kappa}_{\mu\tau}\vert^2+\vert \widetilde{\kappa}_{\tau\mu}\vert^2\right)
\eea
and  
  \bea
\Gamma(H_{1}\to\tau\tau) =  
 \frac{m_{H_{1}}}{8\pi}\left[\cos\theta\; \frac{Y_{\tau}^{\rm diag}}{\sqrt{2}}+\sin\theta \; \widetilde{\kappa}_{\tau\tau}\right]^{2} \;,
\eea
and analogously for $H_1 \rightarrow \mu\mu$. We see that, in our convention, negative $\theta$ reduces  $\Gamma(H_{1}\to\tau\tau)$ without affecting the LFV rates.

The LHC experimental    bounds on Higgs decays into leptons assume that the Higgs production
cross section is not modified by new physics. This is not the case in our model since
both the  $H_1$ production cross section and its total width are reduced by the factor $\cos^2\theta$. Hence, the  experimental limits in fact constrain the combination $\sigma(H_1)\; {\rm BR }(H_1 \rightarrow l_i l_j) $, which we take into account below.

 The LHC  searches for the Higgs decay into tau's yield \cite{Aad:2015vsa},\cite{Chatrchyan:2014nva}
 \begin{eqnarray}
 &&\sigma (H_1)\; {\rm BR }(H_1 \rightarrow \tau\tau)_{\rm ATLAS} = \left(  1.43^{+0.43}_{-0.37} \right) \times
\sigma (h)\; {\rm BR }(h \rightarrow \tau\tau)_{\rm SM} ~, \nonumber\\
 &&\sigma (H_1)\; {\rm BR }(H_1 \rightarrow \tau\tau)_{\rm CMS} = \left(  0.91 \pm 0.28 \right) \times
 \sigma (h)\;  {\rm BR }(h \rightarrow \tau\tau)_{\rm SM} ~,
 \end{eqnarray} 
 where $\sigma$ is the production cross section and  ${\rm BR }(h \rightarrow \tau\tau)_{\rm SM}=0.063$. Combining these results naively
 gives approximately $ \sigma (H_1)\;{\rm BR }(H_1 \rightarrow \tau\tau)= (1.06 \pm 0.23) \times  \sigma (h)\;  {\rm BR }(h \rightarrow \tau\tau)_{\rm SM} $, which we will use as the ``guideline'' bound. 
 This implies that at 95\% CL the tau coupling can be enhanced by no more than 25\% or so,
 if the production cross section is the same as that in the SM. 
Another important constraint comes from  
 the ATLAS limit on the Higgs decay into muons    \cite{Aad:2014xva},
  \begin{equation}
\sigma (H_1)\; {\rm BR }(H_1 \rightarrow \mu\mu) < 1.5 \times 10^{-3} \; \sigma(h)~,
 \end{equation}
 whereas the SM prediction is for ${\rm BR }(h \rightarrow \mu\mu)$ is $2\times 10^{-4}$. This allows for the Higgs--muon  coupling enhancement by a factor of 2.6 or so, again assuming the SM production cross section. 
 
To incorporate the difference between the $H_1$ and $h$ production cross sections, we find it convenient to introduce the effective branching ratio 
${\rm BR_{eff} }(H_1 \rightarrow l_i l_j)$ through 
 \begin{equation}
 \sigma(H_1)\; {\rm BR }(H_1 \rightarrow l_i l_j) = \sigma(h)\;  { \Gamma(H_1 \rightarrow l_i l_j)  \over  \Gamma^{\rm total}_{\rm SM} (h) } \equiv \sigma(h)\; {\rm BR_{eff} }(H_1 \rightarrow l_i l_j) \;,
 \end{equation}
  where the first equality holds up to percent--level corrections
  and $\Gamma^{\rm total}_{\rm SM} (h)=4.1$ MeV.
 The LHC result (\ref{h-tau-mu}) then applies to this effective branching ratio.

 \begin{figure}[ht!]
\begin{center}
\includegraphics[scale=0.72]{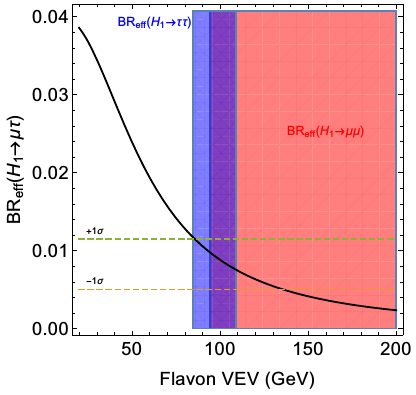}
\end{center}
\caption{  BR$_{\rm eff} (H_1 \to \mu \tau) $ vs $v_\phi$ (black curve)  for  $\sin\theta=-0.3$. The red region is allowed by  ${\rm BR_{eff}  }(H_1 \rightarrow \mu\mu)$ at 95\% CL, the blue region is allowed  by ${\rm BR_{eff}  }(H_1 \rightarrow \tau\tau)$, while their overlap (purple) is consistent with both. The dashed lines show the $\pm 1 \sigma$ limits on the observed BR$_{\rm eff} (H_1 \to \mu \tau)$.
  }
\label{2-LHC}
\end{figure}

 Figure \ref{2-LHC} shows that all of the constraints can be satisfied and the observed 
 BR$_{\rm eff} (H_{1}\to\mu\tau)$  accommodated  for $\sin \theta=-0.3$. This limits the flavon VEV to be around 100 GeV, where partial cancellations between the SM Yukawa coupling and the $\widetilde \kappa$--contribution to  $H_{1}\to\tau\tau$ are effective.\footnote{Large values $v_\phi >700$ GeV are also allowed by BR$_{\rm eff} (H_{1}\to\tau\tau)$. These however lead to negligible   BR$_{\rm eff} (H_{1}\to\mu\tau)$.} The maximal allowed BR$_{\rm eff} (H_{1}\to\mu\tau)$  is close to  1\% for this example.
 Note that BR$_{\rm eff} (H_1 \to l_i l_j)$ is independent of $m_{\Im\phi}$
 and $m_{H_2}$  to leading order, so the latter can be adjusted in order to make
 a particular value of $v_\phi$ consistent with the LFV constraints.

The range of allowed $\vert \sin \theta\vert $ is limited by two factors: values substantially above 0.3 are inconsistent with the latest Higgs coupling data \cite{Higgs-data}
according to which 
\begin{equation}
\vert \sin\theta \vert < 0.33 \;.
\end{equation}
At large $m_{H_2}$ this bound is superseded by that from the electroweak precision measurements  \cite{Falkowski:2015iwa}. 
Values of $\vert \sin \theta\vert $ below 0.2 would require a low new physics scale 
 $\Lambda \sim v_\phi / \epsilon $ in order to accommodate the observed 
 BR$_{\rm eff} (H_{1}\to\mu\tau)$ (Fig.~\ref{1-LHC}, right). For instance, at $\sin \theta =-0.2$, further new physics states 
 are expected to appear at $\Lambda \sim$ 700 GeV, whereas for $\sin \theta =-0.1$ it becomes
 as low as 300 GeV.  Whether such scenarios can be considered realistic depends on the details of the UV completion. While  the flavor physics and collider  bounds are 
 highly model dependent, constraints on the
 multiplicity of states with electroweak quantum numbers are rather weak  \cite{Gross:2016ioi}. All in all, here we make the   assumption that   $\Lambda $
 around 1 TeV can be consistent with the data  in some classes of UV completions.

 One should keep in mind the limitations of the present  approach.   Our effective Froggatt--Nielsen  theory includes only Im$\phi$ and $H_2$ as additional active degrees of freedom. Concrete UV completions would involve further states which can affect our considerations, in particular the loop processes. Hence the LFV bounds we obtain should be treated as 
 ``guidelines''. Also, within the effective theory one cannot explain why  $m_{\Im\phi}$
 is comparable to $v_\phi$, whereas one would naively expect it to be substantially lighter. This issue can presumably be addressed in more sophisticated UV completions,
 where $m_{\Im\phi}$ is generated through a flavor--blind field. 
 
  Nevertheless, we find it encouraging that our simple framework can accommodate all the constraints and fit the observed BR$_{\rm eff} (H_{1}\to\mu\tau)$. The key ingredients are a texture with a large $\mu-\tau$ mixing and a  leptophilic flavon with an electroweak size VEV.
Surprisingly, such a set--up is rather poorly constrained, especially what concerns properties of  Im$\phi$. Since it does not mix with the SM Higgs and couples only to leptons, the best limits would presumably come from exotic Z decays into 4 $\tau$'s.
However, the rate is suppressed by the tau Yukawa coupling squared which makes it too small to place a useful bound on  $m_{\Im\phi}$.

\section{Conclusion}
\label{conclusion}
 
Motivated by the tentative observation of the $h \rightarrow \mu\tau$ decay at the LHC, we have explored a lepton--specific Froggatt--Nielsen  framework   which naturally leads to lepton flavor violation at the observable level.  The corresponding flavon  mixes
with the Standard Model Higgs such that the resulting Higgs--like boson decays to 
$ \mu\tau$ with the branching ratio at the percent level.

This scenario necessitates a flavon VEV at the electroweak scale which we find to be consistent with the LFV  and Higgs data constraints. The  Froggatt--Nielsen symmetry must   be either discrete or softly broken to allow for a massive  Im$\phi$. Due to its pseudoscalar nature, the  latter facilitates substantial cancellations in LFV processes 
and is only weakly constrained by collider  data. 
 
In this work, we have focused on a specific Yukawa texture resulting in  a large mixing in the $\mu$--$\tau$ sector. Further possible charge assignments as well as correlations among 
  observables will be analyzed in our subsequent publication.

\vspace{0.5cm}

{\bf Acknowledgements.}  
KH and VK acknowledge the H2020-MSCA-RICE-2014 grant no. 645722 (NonMinimalHiggs). NK is supported by Vilho, Yrj{\"o} and Kalle V{\"a}is{\"a}l{\"a}
Foundation. OL and VK's work was partially supported by the Academy of Finland project ``The Higgs Boson and the Cosmos'', project no. 267842.

\newpage

\appendix
\section{Exact Yukawa matrix}
\label{detail-kappa}

The eigenvalues of the Yukawa matrix are sensitive to the exact values of the proportionality coefficients, which we provide below.
\begin{equation}
\label{exact-texture}
Y =\left(
\begin{array}{ccc}
 3.3855~ \epsilon ^6 & -0.625~ \epsilon ^6 & 3.5~ \epsilon ^7 \\
 5.36~ \epsilon ^4 & 6.1465~ \epsilon ^4 & -3.125~ \epsilon ^5 \\
 0.5 ~\epsilon ^2  & 0.5 ~\epsilon ^2   & 7.3312~ \epsilon ^3  \\
\end{array}
\right),
\end{equation}
where $\epsilon=0.1$. One can verify that this matrix reproduces the observed lepton masses. 
 It is diagonalized by the unitary transformations $U_L$ and $U_R$ as
\begin{equation}
Y_{\textrm{diag}}=U_L Y U_R^{\dagger},
\end{equation}
with
\begin{equation}
U_{L} \approx\left(
\begin{array}{ccc}
 1 & -1\times 10^{-3} & -8\times 10^{-5} \\
 -1\times 10^{-3} & -1 & 5\times 10^{-2} \\
 -2\times 10^{-4}  & -5\times 10^{-2}  & -1  \\
\end{array}
\right)
 ~~ \textrm{and} ~~
 U_{R} \approx
\left(
\begin{array}{ccc}
0.8  & -0.7  & -6\times 10^{-2}  \\
-0.4  & -0.6  & 0.7  \\
-0.5 & -0.5 & -0.7 \nonumber
\end{array}
\right).
\end{equation}
We see that while $U_L$ is approximately diagonal, $U_R$ involves large angle rotations and is of ``democratic'' form.  This is the key ingredient in obtaining a significant
BR$(H_{1}\to\mu\tau)$.


\begin{thebibliography}{99}
\bibitem{Froggatt:1978nt} 
  C.~D.~Froggatt and H.~B.~Nielsen,
  Nucl.\ Phys.\ B {\bf 147}, 277 (1979).

\bibitem{Khachatryan:2015kon} 
  V.~Khachatryan {\it et al.} [CMS Collaboration],
  Phys.\ Lett.\ B {\bf 749}, 337 (2015)
  [arXiv:1502.07400 [hep-ex]].

\bibitem{Aad:2015gha} 
  G.~Aad {\it et al.} [ATLAS Collaboration],
  arXiv:1508.03372 [hep-ex].

\bibitem{Patt:2006fw} 
V.~Silveira and A.~Zee,
  Phys.\ Lett.\ B {\bf 161}, 136 (1985);   
  R.~Schabinger and J.~D.~Wells,
  Phys.\ Rev.\ D {\bf 72}, 093007 (2005)
  [hep-ph/0509209];
  B.~Patt and F.~Wilczek,
  hep-ph/0605188.


\bibitem{Babu:1999me} 
  K.~S.~Babu and S.~Nandi,
  Phys.\ Rev.\ D {\bf 62}, 033002 (2000)   
  [hep-ph/9907213].

\bibitem{Giudice:2008uua} 
  G.~F.~Giudice and O.~Lebedev,
  Phys.\ Lett.\ B {\bf 665}, 79 (2008)
  [arXiv:0804.1753 [hep-ph]].

\bibitem{Bauer:2015fxa} 
  M.~Bauer, M.~Carena and K.~Gemmler,
  JHEP {\bf 1511}, 016 (2015)
  [arXiv:1506.01719 [hep-ph]].





\bibitem{Goudelis:2011un} 
  A.~Goudelis, O.~Lebedev and J.~h.~Park,
  Phys.\ Lett.\ B {\bf 707}, 369 (2012) 
  [arXiv:1111.1715 [hep-ph]].


\bibitem{Blankenburg:2012ex} 
  G.~Blankenburg, J.~Ellis and G.~Isidori,
  Phys.\ Lett.\ B {\bf 712}, 386 (2012)
  [arXiv:1202.5704 [hep-ph]].






\bibitem{McKeen:2012av} 
  D.~McKeen, M.~Pospelov and A.~Ritz,
  Phys.\ Rev.\ D {\bf 86}, 113004 (2012)
  [arXiv:1208.4597 [hep-ph]].


\bibitem{Davidson:2012ds} 
  S.~Davidson and P.~Verdier,
  Phys.\ Rev.\ D {\bf 86}, 111701 (2012)
  [arXiv:1211.1248 [hep-ph]].


\bibitem{Dery:2013rta} 
  A.~Dery, A.~Efrati, Y.~Hochberg and Y.~Nir,
  JHEP {\bf 1305}, 039 (2013)
  [arXiv:1302.3229 [hep-ph]].


\bibitem{Arroyo:2013tna} 
  M.~Arroyo, J.~L.~Diaz-Cruz, E.~Diaz and J.~A.~Orduz-Ducuara,
  arXiv:1306.2343 [hep-ph].


\bibitem{Crivellin:2013hpa} 
  A.~Crivellin, S.~Najjari and J.~Rosiek,
  JHEP {\bf 1404}, 167 (2014)
  [arXiv:1312.0634 [hep-ph]].


\bibitem{Crivellin:2014cta} 
  A.~Crivellin, M.~Hoferichter and M.~Procura,
  Phys.\ Rev.\ D {\bf 89}, 093024 (2014)
  [arXiv:1404.7134 [hep-ph]].


\bibitem{Kopp:2014rva} 
  J.~Kopp and M.~Nardecchia,
  JHEP {\bf 1410}, 156 (2014)
  [arXiv:1406.5303 [hep-ph]].


\bibitem{Dery:2014kxa} 
  A.~Dery, A.~Efrati, Y.~Nir, Y.~Soreq and V.~Susic,
  Phys.\ Rev.\ D {\bf 90}, 115022 (2014)
  [arXiv:1408.1371 [hep-ph]].




\bibitem{Sierra:2014nqa} 
  D.~Aristizabal Sierra and A.~Vicente,
  Phys.\ Rev.\ D {\bf 90}, no. 11, 115004 (2014)
  [arXiv:1409.7690 [hep-ph]].


\bibitem{deLima:2015pqa} 
  L.~de Lima, C.~S.~Machado, R.~D.~Matheus and L.~A.~F.~do Prado,
  JHEP {\bf 1511}, 074 (2015)
  [arXiv:1501.06923 [hep-ph]].


\bibitem{Dorsner:2015mja} 
  I.~Doršner, S.~Fajfer, A.~Greljo, J.~F.~Kamenik, N.~Košnik and I.~Nišandžic,
  JHEP {\bf 1506}, 108 (2015)
  [arXiv:1502.07784 [hep-ph]].


\bibitem{Goto:2015iha} 
  T.~Goto, R.~Kitano and S.~Mori,
  Phys.\ Rev.\ D {\bf 92}, 075021 (2015)
  [arXiv:1507.03234 [hep-ph]].


\bibitem{He:2015rqa} 
  X.~G.~He, J.~Tandean and Y.~J.~Zheng,
  JHEP {\bf 1509}, 093 (2015)
  [arXiv:1507.02673 [hep-ph]].



\bibitem{Altmannshofer:2015esa} 
  W.~Altmannshofer, S.~Gori, A.~L.~Kagan, L.~Silvestrini and J.~Zupan,
  Phys.\ Rev.\ D {\bf 93}, no. 3, 031301 (2016)
  [arXiv:1507.07927 [hep-ph]].


\bibitem{Feruglio:2015gka} 
  F.~Feruglio, P.~Paradisi and A.~Pattori,
  Eur.\ Phys.\ J.\ C {\bf 75}, no. 12, 579 (2015)
  [arXiv:1509.03241 [hep-ph]].



\bibitem{Bizot:2015qqo} 
  N.~Bizot, S.~Davidson, M.~Frigerio and J.-L.~Kneur,
  arXiv:1512.08508 [hep-ph].



\bibitem{Buschmann:2016uzg} 
  M.~Buschmann, J.~Kopp, J.~Liu and X.~P.~Wang,
  arXiv:1601.02616 [hep-ph].



\bibitem{Botella:2016krk} 
  F.~J.~Botella, G.~C.~Branco, M.~N.~Rebelo and J.~I.~Silva-Marcos,
  arXiv:1602.08011 [hep-ph].


\bibitem{Belusca-Maito:2016axk} 
  H.~Bélusca-Maïto and A.~Falkowski,
  arXiv:1602.02645 [hep-ph].



\bibitem{Heeck:2014qea} 
  J.~Heeck, M.~Holthausen, W.~Rodejohann and Y.~Shimizu,
  Nucl.\ Phys.\ B {\bf 896}, 281 (2015)
  [arXiv:1412.3671 [hep-ph]].
  
  
\bibitem{Banerjee:2016foh} 
  S.~Banerjee, B.~Bhattacherjee, M.~Mitra and M.~Spannowsky,
  arXiv:1603.05952 [hep-ph].
  
  
\bibitem{Alvarado:2016par} 
  C.~Alvarado, R.~M.~Capdevilla, A.~Delgado and A.~Martin,
  arXiv:1602.08506 [hep-ph].
  









\bibitem{Buchmuller:2007zd} 
  W.~Buchmuller, K.~Hamaguchi, O.~Lebedev, S.~Ramos-Sanchez and M.~Ratz,
  Phys.\ Rev.\ Lett.\  {\bf 99}, 021601 (2007)
  [hep-ph/0703078 [hep-ph]].


\bibitem{Lebedev:2011aq} 
  O.~Lebedev and H.~M.~Lee,
  Eur.\ Phys.\ J.\ C {\bf 71}, 1821 (2011)
  [arXiv:1105.2284 [hep-ph]].



\bibitem{Falkowski:2015iwa} 
  A.~Falkowski, C.~Gross and O.~Lebedev,
  JHEP {\bf 1505}, 057 (2015)
  [arXiv:1502.01361 [hep-ph]].
  
  
\bibitem{Tsumura:2009yf} 
  K.~Tsumura and L.~Velasco-Sevilla,
  Phys.\ Rev.\ D {\bf 81}, 036012 (2010)
  [arXiv:0911.2149 [hep-ph]].
  
  
  
  
  
\bibitem{Bellgardt:1987du} 
  U.~Bellgardt {\it et al.} [SINDRUM Collaboration],
  Nucl.\ Phys.\ B {\bf 299}, 1 (1988).

\bibitem{Amhis:2012bh} 
  Y.~Amhis {\it et al.} [Heavy Flavor Averaging Group Collaboration],
  arXiv:1207.1158 [hep-ex].

\bibitem{Hayasaka:2010np} 
  K.~Hayasaka {\it et al.},
  Phys.\ Lett.\ B {\bf 687}, 139 (2010)
  [arXiv:1001.3221 [hep-ex]].

\bibitem{Adam:2013mnn} 
  J.~Adam {\it et al.} [MEG Collaboration],
  Phys.\ Rev.\ Lett.\  {\bf 110}, 201801 (2013)
  [arXiv:1303.0754 [hep-ex]].



\bibitem{Bertl:2006up} 
  W.~H.~Bertl {\it et al.} [SINDRUM II Collaboration],
  Eur.\ Phys.\ J.\ C {\bf 47}, 337 (2006).






\bibitem{Harnik:2012pb} 
  R.~Harnik, J.~Kopp and J.~Zupan,
  JHEP {\bf 1303}, 026 (2013)
  [arXiv:1209.1397 [hep-ph]].




\bibitem{Kitano:2002mt} 
  R.~Kitano, M.~Koike and Y.~Okada,
  Phys.\ Rev.\ D {\bf 66}, 096002 (2002)
  [Phys.\ Rev.\ D {\bf 76}, 059902 (2007)]
  [hep-ph/0203110].

\bibitem{Alonso:2012ji} 
  R.~Alonso, M.~Dhen, M.~B.~Gavela and T.~Hambye,
  JHEP {\bf 1301}, 118 (2013)
  [arXiv:1209.2679 [hep-ph]].



\bibitem{Barr:1990vd} 
  S.~M.~Barr and A.~Zee,
  Phys.\ Rev.\ Lett.\  {\bf 65}, 21 (1990)
  Erratum: [Phys.\ Rev.\ Lett.\  {\bf 65}, 2920 (1990)].





\bibitem{Aad:2015vsa}
G.~Aad {\it et al.} [ATLAS Collaboration],
JHEP {\bf 1504}, 117 (2015)
[arXiv:1501.04943 [hep-ex]].

\bibitem{Chatrchyan:2014nva} 
V.~Khachatryan {\it et al.} [CMS Collaboration],
  Eur.\ Phys.\ J.\ C {\bf 75}, no. 5, 212 (2015)
 [arXiv:1412.8662 [hep-ex]];
  S.~Chatrchyan {\it et al.} [CMS Collaboration],
  JHEP {\bf 1405}, 104 (2014)
  [arXiv:1401.5041 [hep-ex]].

 



\bibitem{Aad:2014xva}
  G.~Aad {\it et al.} [ATLAS Collaboration],
  Phys.\ Lett.\ B {\bf 738}, 68 (2014)
  [arXiv:1406.7663 [hep-ex]].


\bibitem{Higgs-data} 
  The ATLAS and CMS Collaborations,
  ATLAS-CONF-2015-044.



\bibitem{Gross:2016ioi} 
  C.~Gross, O.~Lebedev and J.~M.~No,
  arXiv:1602.03877 [hep-ph].

















 

\end{thebibliography}
\end{document}